# Streamlining the Selection Phase of Systematic Literature Reviews (SLRs) Using AI-Enabled GPT-4 Assistant API

Seyed Mohammad Ali Jafari

*PhD candidate, Technological Entrepreneurship Department, Faculty of Entrepreneurship, University of Tehran, Iran*

sma_jafari@ut.ac.ir

**Abstract:**

The escalating volume of academic literature presents a formidable challenge in staying updated with the newest research developments. Addressing this, this study introduces a pioneering AI-based tool, configured specifically to streamline the efficiency of the article selection phase in Systematic Literature Reviews (SLRs). Utilizing the robust capabilities of OpenAI's GPT-4 Assistant API, the tool successfully homogenizes the article selection process across a broad array of academic disciplines. Implemented through a tripartite approach consisting of data preparation, AI-mediated article assessment, and structured result presentation, this tool significantly accelerates the time-consuming task of literature reviews. Importantly, this tool could be highly beneficial in fields such as management and economics, where the SLR process involves substantial human judgment. The adoption of a standard GPT model can substantially reduce potential biases and enhance the speed and precision of the SLR selection phase. This not only amplifies researcher productivity and accuracy but also denotes a considerable stride forward in the way academic research is conducted amidst the surging body of scholarly publications.

## 1. Introduction:

In numerous academic disciplines, the rapid surge of articles being published is a reality that presents considerable challenges. This escalation, noted by Ferreira, Fernandes, & Kraus, 2019; Mora, Deakin, & Reid, 2019; Mustak, Salminen, Plé, & Wirtz, 2021, puts significant pressure on researchers conducting systematic literature reviews (SLRs). Today, with so much data available, it's more important than ever to correctly identify, evaluate, and summarize all relevant literature. Traditional methods, however, have difficulty keeping up with the sheer volume of information. Due to the wealth of knowledge available, we need more effective approaches. The traditional SLR process can be quite time-consuming (Hiebl, 2023). Additionally, it increases the potential for subjective bias and oversight, as described by Tomassetti, Rizzo, Vetro, Ardito, Torchiano, & Morisio, 2011, compromising the comprehensiveness expected of SLRs.

Fortunately, we can alleviate some of these shortcomings with advanced NLP-based tools. To bridge this gap, I've developed an AI tool that incorporates OpenAI's Assistant API for automating the analysis selection stage of SLRs. The tool is versatile, applicable to all disciplines, and aims to elevate scholarly work that depends on comprehensive literature reviews. To demonstrate its benefits, I'll provide an example within the field of business and management studies. In the "Implication" section, you'll see how this tool improves the SLR process and offers substantial contributions to academic research tools.

## 2. Background:

### 2.1. Generative Pre-trained Transformers (GPTs) and Their Scope:

Pioneered by OpenAI, Generative Pre-trained Transformers (GPTs) symbolize a remarkable milestone in the realm of large language models (LLMs). Trained exhaustively on massive data sampled from a wide variety of internet sources, these models demand considerable computational resources. The arrival of ChatGPT in 2022, which provided a user-friendly platform for users to engage with the GPT-3.5 model, showcased the immense potential these models embody. Scholars from various fields have acknowledged the prowess of subsequent advancements, such as the GPT-4 model, as indicated by Davidsson and Sufian (2023). These models are capable of conducting intricate analysis that outmatches human cognitive abilities, especially in areas typically dependent on human discretion, like management and entrepreneurship. GPT-4 has particularly been hailed as a superior tool for data analysis, thereby augmenting research methodologies with unparalleled precision levels. For the purposes of this article, the gpt4 model was selected for interaction with the Assistant API.

*2.2. Introduction to OpenAI's Assistant API:*

At its core, an Application Programming Interface (API) is a suite of tools and protocols enabling seamless interaction between different software components. APIs simplify the communication between independently built codebases and supporting services, preempting the need for developers to construct complex functions from the ground up. Therefore, utilizing APIs allows developers to channel their efforts towards the integral aspects of their application, while concurrently utilizing external services to add supplementary functionalities (Sill, 2015).

In this realm, OpenAI's Assistant API gives developers direct access to the vast capabilities of GPT models. It provides a programmatic conduit for product development beyond the regular ChatGPT interface. Designed to automate the integration of advanced natural language processing capabilities into various application contexts (OpenAI, 2023).

The prime factor for choosing the Assistant API over the classical ChatGPT interface lies in the former's enhanced operational proficiency and scalability. Manually analyzing extensive datasets through ChatGPT is not feasible due to the time-intensive nature of the task and the subsequent unstructured data output. Conversely, the Assistant API allows developers to script codes that automate these interactions, enabling direct entry of texts into analytical procedures and efficient extraction of processed data. Such a capability is significantly beneficial when conducting systematic data evaluation, a common feature of SLRs and comparable research undertakings.

*2.3. Systematic Literature Reviews (SLRs) demystified:*

Systematic Literature Reviews (SLRs) acquire their distinctive appeal from their rigorously organized approach towards the collection, evaluation, and synthesis of all pertinent research on a specified subject. The ultimate goal is to assemble an exhaustive review of existing literature (Tranfield et al., 2003).

A case in point is the research conducted by Giuggioli and Pellegrini (2023), who set out on a systematic quest to uncover the nexus between artificial intelligence (AI) and entrepreneurship. Their intricate SLR process entailed formulating a research query, establishing a research protocol, conducting detailed searches across multiple databases, selecting and excluding studies, extracting and analyzing data, and synthesizing findings.

In this article, our aim is to describe an automation tool, powered by the Assistant API and implemented in Python. Its design is intended to ease the researcher's task specifically during the selection phase of a Systematic Literature Review (SLR), which involves selecting and excluding studies. The subsequent sections detail the step-by-step procedure researchers can follow to leverage this tool, which is powered by GPT models via the Assistant API. While various GPT models could be chosen, the gpt-4 model was selected for usage in developing this tool to enhance the identification of relevant studies. Hypothetically, had Giuggioli and Pellegrini (2023) utilized this code, they could have significantly expedited their selection process.

**3. Design:**

*3.1. Set-Up Phase:*

Drawing from Giuggioli and Pellegrini (2023), the SLR process incorporates various steps. However, considering the specific focus of this article, we have named all the steps preceding the process of selecting and excluding studies as the 'set-up phase'. These initial stages include formulating a research query, establishing a research protocol, and conducting exhaustive searches across several databases to gather a comprehensive dataset. This 'set-up phase' as we have termed it, precedes the actual data collection. It necessitates researchers to download a comprehensive dataset by conducting query-driven searches across several databases, including but not limited to, Web of Science (WoS), Scopus, and EBSCO.

*3.2. Stage One:*

The initial task includes identifying and eliminating any duplicate records across datasets retrieved from different databases. Researchers must be certain that none of the essential columns ("Authors", "Article Title", "Abstract") in each article is empty to enhance the process of duplicate removal. Consistency must be ensured in naming the three identifier columns ("Authors", "Article Title", "DOI") across all source Excel files. The code outlined in Figure 1 can then be used to merge all files into a single dataset, simultaneous eradicating duplicates.

```python
import pandas as pd
import os

def read_and_combine_excel_files(file_paths):
    all_data = pd.DataFrame()
    for file_path in file_paths:
        df = pd.read_excel(file_path)
        all_data = all_data.append(df, ignore_index=True)
    return all_data

def remove_empty_entries(data):
    required_fields = ["Authors", "Article Title", "Abstract"]
    # Count entries before removal
    before_removal = len(data)
    # Remove entries with any empty required fields
    data = data.dropna(subset=required_fields, how='any').reset_index(drop=True)
    # Count entries after removal
    after_removal = len(data)
    # Calculate and return the number of removed entries because of being empty
    removed_because_empty = before_removal - after_removal
    return data, removed_because_empty

def remove_duplicates(data):
    before_removal = len(data)
    data = data.drop_duplicates(subset='DOI').reset_index(drop=True)
    data_no_doi = data[data['DOI'].isna() | (data['DOI'] == '')]
    data_doi_present = data[~data['DOI'].isna() & (data['DOI'] != '')]
    cols_to_check = ['Authors', 'Article Title']
    data_no_doi = data_no_doi.drop_duplicates(subset=cols_to_check).reset_index(drop=True)
    cleaned_data = pd.concat([data_doi_present, data_no_doi], ignore_index=True)
    after_removal = len(cleaned_data)
    return cleaned_data, before_removal, before_removal - after_removal

# List all file paths
file_paths = ['/content/Data_scopus.xls', '/content/Data_wos.xls']  # Replace with actual paths

# Read and combine all Excel files
combined_data = read_and_combine_excel_files(file_paths)

# Remove entries with empty fields
combined_data, removed_empty = remove_empty_entries(combined_data)
print(f"Total articles removed due to empty fields: {removed_empty}")

# Remove duplicates and get the cleaned data
cleaned_data, total_articles, removed_duplicates = remove_duplicates(combined_data)

# Print the results
print(f"Total articles processed: {total_articles}")
print(f"Total duplicates removed: {removed_duplicates}")
print(f"Total articles removed due to empty fields: {removed_empty}")

# save the cleaned data to a new Excel file.
cleaned_data.to_excel('/content/cleaned_articles.xlsx', index=False)
```

Figure 1 - Duplicate Removal Code Python script that identifies and removes duplicate articles in the dataset

*3.3. Stage Two:*

Before proceeding to this stage, it's essential to install the OpenAI package using the command (!pip install openai --use-deprecated=legacy-resolver). Following this, researchers can implement the code highlighted in Figure 2. The Assistant API is then employed to evaluate the eligibility of each article for inclusion in the study. Additional information, including authors' names, article titles, and publication years, is also included to ensure a structured and clear output. Researchers provide the Assistant API with instructions on how to determine article acceptance or rejection based on specific criteria relating to the Systematic Literature Review (SLR). As presented in Figure 3, I wrote an instruction that the Assistant API could use to guide the GPT model. This instruction helps GPT-4 understand what it should do and how the output should look. For example, in this instruction, I asked the Assistant to have the GPT model format its selection decision to include 'Acceptance' (indicating inclusion in the study), 'Methodology' (categorizing the article as theoretical, empirical quantitative, or qualitative), and 'Explanation' (providing insights into the API's decision-making process). These results further include author names, article titles, and publication years previously inputted by the researcher for consistent output.

```python
import pandas as pd
from openai import OpenAI
import os
os.environ['OPENAI_API_KEY'] = '***'

client = OpenAI()
all_responses = []  # List to store all messages
threads=[]

def read_excel(file_path):
    # Read the excel file
    df = pd.read_excel(file_path)
    # Adjust this to include additional columns as needed
    return df[['Abstract', 'Authors', 'Article Title', 'Publication Year']]

def evaluate_article(abstract, authors, title, year):
    thread = client.beta.threads.create()
    threads.append(thread)
    # Send the abstract, authors, and additional data to the assistant
    content = f"Abstract: {abstract}\nAuthors: {authors}\nArticle Title: {title}\nPublication Year: {year}"
    message = client.beta.threads.messages.create(
        thread_id=thread.id,
        role="user",
        content=content
    )

    # Run the assistant and retrieve the run result
    run = client.beta.threads.runs.create(thread_id=thread.id, assistant_id="asst_iPZzJp2vcu434bYjuYfXRiBv")
    run = client.beta.threads.runs.retrieve(thread_id=thread.id, run_id=run.id)

    # List messages for analysis
    messages = client.beta.threads.messages.list(thread_id=thread.id)
    return messages

def main():
    file_path = '/content/MainData.xlsx'
    articles = read_excel(file_path)
    all_responses = []  # Initialize a list to store all responses

    for index, row in articles.iterrows():
        response = evaluate_article(row['Abstract'], row['Authors'], row['Article Title'], row['Publication Year'])
        all_responses.append(response)  # Store the response in the list

    # Process and print the responses
    for response in all_responses:
        print(response)

if __name__ == "__main__":
    main()
```

Figure 2 - Article Inclusion Code Python code using Assistant API to decide if an article should be included in the review

Your primary function is to analyze academic articles related to the impact of AI on entrepreneurial decision-making. You must evaluate each article's relevance and suitability for inclusion in a systematic literature review (SLR). Consider the following aspects in your evaluation:

1. Relevance to the Topic: Assess if the article's content is directly related to the use of AI in entrepreneurial decision-making. Exclude articles that do not focus on this intersection.

2. Abstract Analysis: Analyze the abstract of each article for key insights, methodologies, and findings that contribute to understanding the impact of AI on entrepreneurial decisions.

Your analysis should conclude with a clear recommendation on whether the article should be included in the SLR for further analysis. Provide a very brief justification for your decision based on the criteria above.

your style of output should be like the following:
Acceptance:(if the article is acceptable or not) Yes/No
Authors:
Article Title:
Publication Year:
Methodology:(tell about it's methodology if it's (theoretical paper- empirical (quantitative)-empirical (qualitative)))
Explanation:
don't add any \n or anything else except raw text

Figure 3 - Decision-Making Guide for Assistant API A graphic representation showing the criteria used by the Assistant API in deciding article inclusion.

*3.4. Stage Three:*

Before embarking on the final stage, the pytesseract package needs to be installed (!pip install pytesseract). Subsequently, the code depicted in Figure 4 compiles all assistant API responses and generates an Excel file as the output. This file systematically collates the articles selected for further analysis and synthesis by the researcher as part of the SLR process.

```python
import pandas as pd
from PIL import Image
import pytesseract

def parse_content(content):
    # Initialize defaults
    acceptance = ''
    author_names = ''
    article_title = ''
    publication_year = ''
    methodology = ''
    explanation = ''

    # Split the content into parts
    parts = content.split('\n')

    # Parse each line for specific keys
    for part in parts:
        if part.startswith("Acceptance: "):
            acceptance = part.split('Acceptance: ')[1]
        elif part.startswith("Authors: "):
            author_names = part.split('Authors: ')[1]
        elif part.startswith("Article Title: "):
            article_title = part.split('Article Title: ')[1]
        elif part.startswith("Publication Year: "):
            publication_year = part.split('Publication Year: ')[1]
        elif part.startswith("Methodology: "):
            methodology = part.split('Methodology: ')[1]
        elif part.startswith("Explanation: "):
            explanation = part.split('Explanation: ')[1]

    return acceptance, author_names, article_title, publication_year, methodology, explanation

def convert_messages_to_dataframe(messages):
    data = []
    for message in messages.data:
        try:
            message_id = getattr(message, 'id', '')
            assistant_id = getattr(message, 'assistant_id', '')
            # Correctly concatenate the content texts
            content = " ".join([getattr(text.text, 'value', '') for text in message.content])
            created_at = getattr(message, 'created_at', '')
            role = getattr(message, 'role', '')
            thread_id = getattr(message, 'thread_id', '')

            # Parse the content
            acceptance, author_names, article_title, publication_year, methodology, explanation = parse_content(content)
            if (role == "assistant"):
                data.append({
                    "Acceptance": acceptance,
                    "Article Title": article_title,
                    "Methodology": methodology,
                    "Explanation": explanation,
                    "Authors": author_names,
                    "Publication Year": publication_year,
                    "Message ID": message_id,
                    "Assistant ID": assistant_id,
                    "Created At": created_at,
                    "Role": role,
                    "Thread ID": thread_id
                })
        except Exception as e:
            print(f"Error processing message: {e}")

    return pd.DataFrame(data)

mainDf = pd.DataFrame()

# Assuming threads is defined and client.beta.threads.messages.list is a function that retrieves messages
for thread in threads:
    messagestemp = client.beta.threads.messages.list(thread_id=thread.id)
    df_temp = convert_messages_to_dataframe(messagestemp)
    mainDf = mainDf.append(df_temp, ignore_index=True)

# Save the main DataFrame to an Excel file
mainDf.to_excel("AnalyzedArticles.xlsx", index=False)
```

Figure 4 - Final Output Creation Code Python code snippet that assembles and structures the API responses into a final Excel file.

**4. Deployed Example:**

The tool introduced in this article employs automation to streamline the Systematic Literature Review (SLR) selection phase, demonstrating its proficiency across various academic disciplines. For practical demonstration, I've utilized this tool during the selection phase of an SLR centered around 'Artificial Intelligence and Entrepreneurship'. The aim was to make use of this tool for the efficient analysis of a wide scope of articles in the domains of business, economics, and management, thereby emphasizing its competency in selecting pertinent articles for review.

*4.1. Set-Up Phase:*

During the 'set-up phase', I identified appropriate keywords to streamline database searches. Subsequently, I downloaded two separate datasets containing article information: one from Scopus and another from Web of Science (WoS).

*4.2. Stage One:*

In the first stage, I utilized these two downloaded datasets from Scopus and WoS. Using the stage-one code, I processed roughly 1,499 articles, identifying 210 duplicates. Additionally, one article was removed due to the lack of crucial information in the 'Authors', 'Article Title', and 'Abstract' sections.

*4.3. Stage Two:*

Transitioning into the second stage, I used the integrated dataset and the stage-two code to analyze which articles warranted inclusion in the study, while also gaining insights into the reasons for potential rejections.

*4.4. Stage Three:*

Lastly, in the third and final stage, the stage-three code was applied to generate an output of this data following complete analyses from the Assistant API. This data was then stored in the form of an Excel file. Figure 5, presenting a snapshot of the finalized analysis results, allows researchers to promptly ascertain which articles should be included in the study. Plus, it helps them comprehend the Assistant API's decision-making process and the unique methodologies of each inspected article.

| Acceptance | Article Title | Methodology | Explanation | Authors | Publication Year |
|---|---|---|---|---|---|
| No | Smart, hybrid and context-aware POI mobile recommender system in tourism in Oman | Theoretical paper | The article's focus is on the development of a mobile recommender system for the tourism sector, optimizing the tourist experience by managing time and budget constraints. Although it involves machine learning algorithms, the study's primary application area is tourism, and it specifically relates to Point of Interest (POI) recommendations. There is a conspicuous absence of discussion about AI's influence on entrepreneurial decision-making. Hence, it does not align directly with the intersection of AI and entrepreneurial decision-making. | Afsahhosseini F.; Al-Mulla Y. | 2023 |
| Yes | Natural Language Processing for Innovation Search – Reviewing an Emerging Non-human Innovation Intermediary | Theoretical paper | This article is highly relevant to the topic as it deals with the application of AI, specifically NLP, in the context of innovation search, which is an integral part of entrepreneurial decision-making. The use of AI to facilitate early-stage innovation is directly related to strategic decisions entrepreneurs must make. The methodology involves a review of 167 academic articles, aiming at a comprehensive understanding of how NLP functions as an innovation intermediary. As it provides insights into how entrepreneurs can leverage AI for innovation search, the article would likely contribute significantly to a systematic literature review on the impact of AI in entrepreneurial decision-making. | Just J. | 2024 |
| No | Global techno-politics: A review of the current status and opportunities for future research | Theoretical paper | The abstract indicates that the article focuses on the broad relationship between technology and global politics, specifically addressing the concept of global techno-politics (GTP). It does not concentrate on the use of artificial intelligence (AI) in entrepreneurial decision-making, but rather on the overarching impact of technology on geopolitical dynamics and policy-making. As the article's content does not directly address the intersection of AI and entrepreneurial decision-making, it is not relevant for inclusion in a systematic literature review focusing specifically on that topic. | Yan J.; Leidner D.E.; Peters U. | 2024 |
| No | Enhancing lifestyle and health monitoring of elderly populations using CSA-TkELM classifier | Empirical (Quantitative) | The abstract indicates that the article's primary focus is on Human Activity Recognition (HAR) for health monitoring in the elderly population using wearable sensor data and an optimized classifier algorithm. Despite the application of intelligent algorithms and technology reflective of AI approaches, the article does not discuss AI within the context of entrepreneurial decision-making. As such, it does not meet the criteria for relevance to the topic as established for the SLR. The article appears to be empirical and quantitative, centering on performance metrics to evaluate the classifier's effectiveness, but this is outside the scope of an SLR on AI's impact on entrepreneurial decision-making. | Rosaline R.A.A.; Ponnuviji N.P.; T.C. S.L.; Manisha G. | 2023 |
| No | How Do Fast-Fashion Copycats Affect the Popularity of Premium Brands? Evidence from Social Media | Empirical (Quantitative) | While the article incorporates advanced analytics, such as deep learning image analytics, and examines an interesting domain of consumer behavior related to fashion brands, it does not directly address the impact of AI on entrepreneurial decision-making. The focus is on the interactions between high-end fashion brands and fast-fashion copycats, with an analysis of consumer posting behaviors on social media, which is tangential to the specific topic of AI's influence on entrepreneurial decisions. Thus, the article should not be included in the SLR seeking to understand the influence of AI in the decision-making process of entrepreneurs. | Shi Z.; Liu X.; Lee D.; Srinivasan K. | 2023 |

Figure 5 - Example of Completed Excel Output - A partial view of the structured, analytically-generated final Excel output file.

## 5. Discussion:

### 5.1. Benefits:

The adoption of GPT-4 through OpenAI's Assistant API ushers in a pivotal progression in automating the selection stage of Systematic Literature Reviews (SLRs). It equips researchers with a potent instrument for efficiently navigating the exponentially increasing sea of academic literature. This method notably trims the time spent on detecting relevant studies, curbs the probability of human error, and imparts a more impartial selection process, underpinned by established criteria.

### 5.2. Limitations:

Despite these benefits, this tool is not devoid of limitations. Its efficiency hinges on the meticulous designation of selection criteria. Moreover, while AI affords objectivity in criterion application, it may inadvertently mirror biases ingrained in its training data. Therefore, human expertise continues to hold crucial value in certifying the pertinence and caliber of selected articles.

### 5.3. Future Research Directions:

To further capitalize on the potential of AI in SLRs, future research could broaden the scope of AI-based services. Developing AI-driven tools for deeper analysis, such as conducting thematic reviews using GPT models under researcher supervision, stands out as a promising direction. These advancements could result in comprehensive AI-assisted tools, encompassing the entire SLR process while preserving rigorous academic standards.

**6. Implications:**

*6.1. Practical Implications:*

From a practical standpoint, the approach unveiled here stands to aid practitioners by assuring swift access to pertinent literature, hence bolstering evidence-informed decision-making. The heightened efficiency in executing SLRs equips professionals to keep abreast of the freshest research insights.

*6.2. Academic Implications:*

From an academic perspective, the automated tool's significance lies in its ability to refine the literature review process. It enables a more strategic deployment of researcher effort, reorienting focus from manual article selection towards critical examination and interpretation. Ultimately, this contributes to enhancing the quality of systematic literature reviews.

**7. Conclusion:**

The development of an AI-based tool, as detailed in this paper, optimizes the selection phase of systematic literature reviews by leveraging the strengths of GPT-4 through the OpenAI's Assistant API. This method represents a practical response to the challenges posed by the expanding corpus of literature, improving both efficiency and accuracy. While it offers improvements to current manual approaches, its full potential is yet to be realized and will likely be reached through ongoing refinement and expansion into subsequent stages of the SLR process. Future iterations could see the role of Assistant API and AI technology expanding to assist with complex analytical tasks such as thematic analysis, all under the careful stewardship of researchers. This tool's continued development has the potential to deeply impact academic research practices, streamlining the management and synthesis of scholarly knowledge in an age characterized by information abundance.